\documentclass[conference,10pt]{IEEEtran}
\IEEEoverridecommandlockouts
\usepackage{cite}
\usepackage{amsmath,amssymb,amsfonts}
\usepackage{graphicx}
\usepackage{subfigure}
\usepackage{textcomp}
\usepackage{xcolor}
\usepackage{soul,color}
\usepackage{bm}
\usepackage{booktabs}
\usepackage{array,multirow,graphicx}
\usepackage{url}
\usepackage{mdwlist}
\usepackage{neuralnetwork}
\usepackage{balance}

\DeclareRobustCommand{\IEEEauthorrefmark}[1]{\smash{\textsuperscript{\footnotesize #1}}}

\begin{document}
\title{Detecting 5G Narrowband Jammers with CNN, $k$-nearest Neighbors, and Support Vector Machines\thanks{This work was partially supported by the German Federal Office for Information Security within the project ADWISOR5G under grant ID 01MO23030B.}}

\author{
\IEEEauthorblockN{Matteo Varotto\IEEEauthorrefmark{1}, Florian Heinrichs\IEEEauthorrefmark{3},  Timo Schürg\IEEEauthorrefmark{1}, Stefano Tomasin\IEEEauthorrefmark{2}, and Stefan Valentin\IEEEauthorrefmark{1}}\smallskip
\IEEEauthorblockA{\url{matteo.varotto@h-da.de}, \url{f.heinrichs@fh-aachen.de}, \url{timo.schuerg@h-da.de}, \url{tomasin@dei.unipd.it}, \url{stefan.valentin@h-da.de}}\medskip
\IEEEauthorblockA{
\IEEEauthorrefmark{1}Dep. of Computer Science, Darmstadt University of Applied Sciences, Germany\\
\IEEEauthorrefmark{2}Dep. of Information Engineering, Dep. of Mathematics, University of Padova, Italy\\
\IEEEauthorrefmark{3}Dep. of Medical Engineering and Technomathematics, FH Aachen -- University of Applied Sciences, Germany
}
}

\maketitle

\begin{abstract}
5G cellular networks are particularly vulnerable against narrowband jammers that target specific control subchannels in the radio signal. One mitigation approach is to detect such jamming attacks with an online observation system, based on machine learning. We propose to detect jamming at the physical layer with a pre-trained machine-learning model that performs binary classification. Based on data from an experimental 5G network, we study the performance of different classification models. A convolutional neural network will be compared to support vector machines and $k$-nearest neighbors, where the last two methods are combined with principal component analysis. The obtained results show substantial differences in terms of classification accuracy and computation time.
\end{abstract}


\section{Introduction}
5G cellular networks promise improved latency, data rate, and the support of mission-critical applications. Consequently, their application in automotive and smart manufacturing is rapidly increasing \cite{5g_iot}.

Unfortunately, 5G radio signals are particularly vulnerable to jamming attacks \cite{jamming_survey}. A jammer performs a denial of service attack by emitting interference in the frequency band used by, e.g., a 5G network. By targeting specific control subchannels in the signal, it is sufficient to interfere only with a small fraction of the overall bandwidth to disrupt the service. Such narrowband jammer can be easily implemented as a software-defined radio (SDR) at low cost. One very effective narrowband jamming approach \cite{ssb_jammer} targets the signal synchronization block (SSB) in the 5G signal \cite{3gpp:5gssb}, which is critical to establishing and maintaining 5G communication. Other narrowband jamming approaches are discussed in \cite{jamming_survey}; in particular in \cite{robinson} it is documented how to locate in the frequency domain narrowband interference, but with no major effects on the reliability of the communication as in our case.

One mitigation approach is to detect jamming attacks by an online wireless intrusion prevention system (WIPS). Such detection often relies on specific parameters such as signal-to-noise-ratio (SNR), bit error rate (BER), packet error rate (PER) \cite{jamm_det_1}, orthogonality conditions\cite{ssb_paper}, or on measurements aggregated over multiple network layers \cite{jamm_det_4}. Such high-level measurements, however, may be easily evaded by a jammer. For instance, a short jamming impulse in the SSB-subband effectively disrupts 5G communication \cite{ssb_jammer} but barely affects the, usually time-averaged, SNR of the main band. The same argument holds for SNR derivatives such as BER and PER. Detection based on measuring the quality of the radio signal at the physical layer \cite{jamm_det_3}, requires accurate synchronization in time and frequency. This not only leads to costly implementation but may also mislead the classification due to normal clock differences in a cellular network.

In this paper, we propose a WIPS to detect 5G narrowband jammers. This detection is based on spectrograms of the 5G radio signal and on a pre-trained machine-learning model for binary classification. The spectrograms are obtained from an experimental 5G network, which is occasionally attacked by an SSB jammer. The objective of our paper is to compare how accurate and how quickly the SSB jammer can be detected with different machine learning models. To this end, we compare the classification accuracy and computational complexity of a convolutional neural network (CNN) with support vector machines (SVMs) and $k$-nearest neighbors (KNNs). SVM and KNN are combined with principal component analysis (PCA) for dimensionality reduction.

The considered security scenario is a typical private 5G network used in industrial applications. We assume at least one available wireless channel for communication, provided by a 5G base station, called gNB. The channel is utilized by the User Equipment (UE) to transmit or receive data whenever need be. The communication channel may be attacked by a jammer in close range of the gNB. 

This network is monitored by a watchdog -- a separate network element placed within range of the gNB. The watchdog receives radio signals and converts them to the digital baseband in the form of in-phase and quadrature (IQ) samples. On these IQ samples, anomaly detection is performed with a pre-trained machine learning model.

This configuration has several benefits. The watchdog can be deployed independently of the cellular network and, as a mere receiver, does not transmit signals that can be detected. Thus, it can be deployed invisibly to a potential attacker. The watchdog is simpler and easier to construct than a communication device since it can still operate effectively with imperfect synchronization in time and frequency. 

The following Sec. \ref{sec:dataset} covers the system assumptions, while in Sec. \ref{sec:method} the machine learning models are explained. In Sec. \ref{sec:results} results are discussed and Sec. \ref{sec:concl} concludes the paper.

\section{System Assumptions}
\label{sec:dataset}
This section will cover the main assumptions, the experimental setup, and the creation of the dataset.

\subsection{Main assumptions}
We focus on a single cell in a mobile network, including the nodes described above. The gNB, one or multiple UEs, a watchdog, and a jammer are all in range of each other. We assume that the watchdog knows the basic radio parameters such as center frequency, bandwidth, and the pilot structure. This assumption is feasible since the required information is constant and known to the operator of a 5G network. 
\newline
For each transmitter-receiver path, the baseband-equivalent received signal per subcarrier $s \in \mathcal{S}$ can be defined as
\begin{equation}
y_s = h_{s} x_s + w_s,
\end{equation}
where $x_s$ is the transmitted orthogonal frequency-division multiplexing (OFDM) symbol, $h_{s}$ is the gain for the channel between the receiver and the transmitter, and $w_s$ is the additive white Gaussian noise (AWGN) at the receiver. All these quantities are complex signals.
\newline
If a jammer starts injecting noise over the same subcarrier, the received signal will become
\begin{equation}
    y_s = h_{s} x_s +\hat{h}_{s} \hat{x}_s + w_s,
\end{equation}
where $\hat{x}_s$ is the signal transmitted by the jammer and $\hat{h}_{s}$ is the gain for the channel between the receiver and the jammer.

For each subcarrier $s$, the watchdog collects an IQ sample, creating a continuous stream of samples from which spectrograms will be created and analyzed to detect the presence of a jammer.

\subsection{Studied Jammer}
We focus on detecting a narrowband jammer that targets the SSB. Unlike traditional jammers, an SSB jammer does not have to interfere with the complete frequency band $\mathcal{S}$ in order to disrupt communication. Instead, it generates interference only on those subcarriers $\mathcal{S}_\textnormal{SSB} \subset \mathcal{S}$ used for synchronization. Since $|\mathcal{S}_\textnormal{SSB}| \ll |\mathcal{S}|$, an SSB jammer uses substantially lower bandwidth and energy than traditional jamming. This improves not only the runtime of battery-powered jammers but also heavily reduces their cost, which is mainly defined by Radio Frequency (RF) bandwidth. The comparably low signal bandwidth also makes SSB jamming more difficult to detect than traditional (full band) jamming. These properties make the SSB jammer a serious threat to 5G networks.

\begin{figure}
    \centering
    \includegraphics[width=1\hsize]{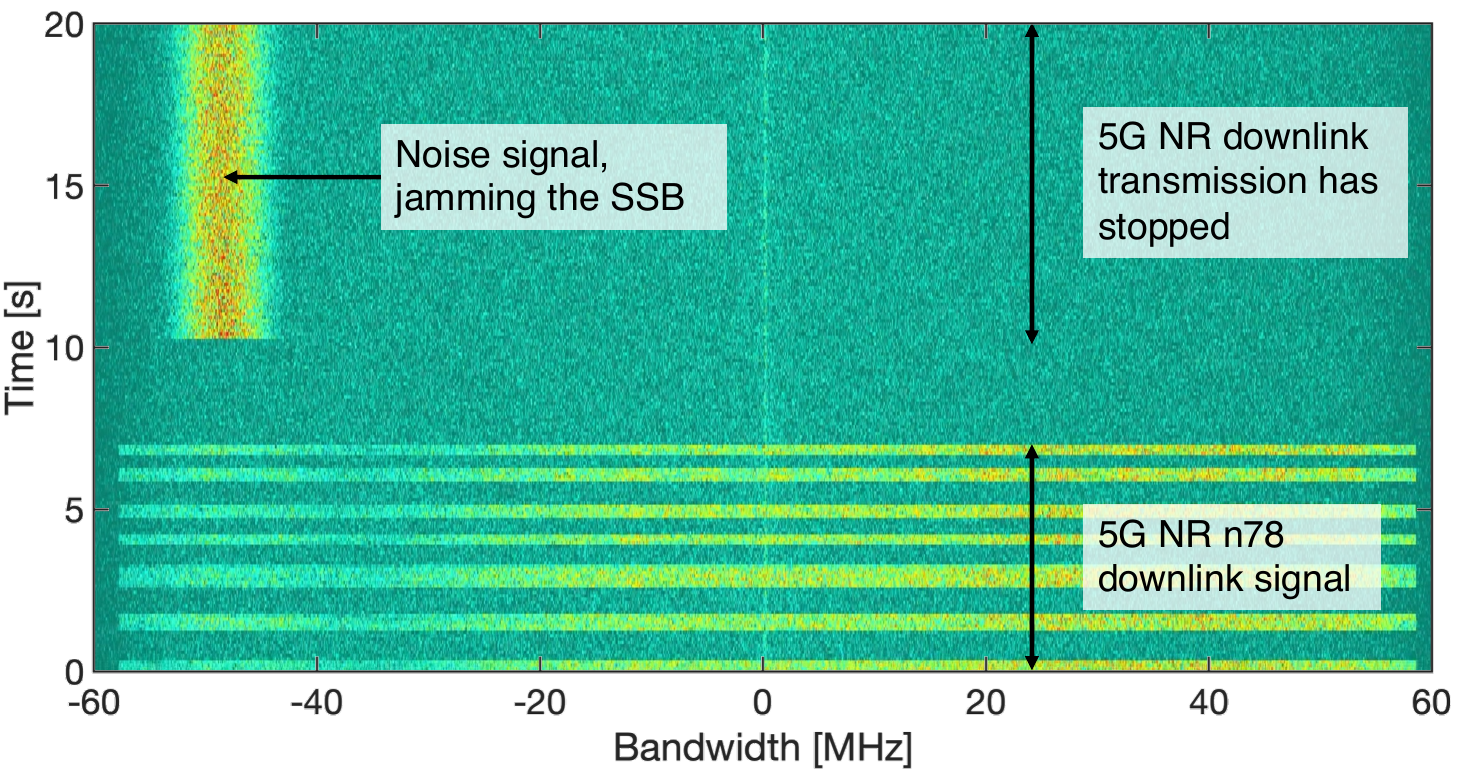}
    \caption{RF spectrogram measured over a bandwidth of $120$ MHz (x-axis) at center frequency $3750$ MHz for 20\,s (y-axis), showing a 5G NR downlink signal of $100$ MHz bandwidth and a signal of an SSB jammer of $7.2$ MHz bandwidth. When the jammer is active, the 5G downlink transmission stops.}
    \label{ssb_spectrum}
\end{figure}

As an example, Fig. \ref{ssb_spectrum} shows a measured spectrogram of a 5G new radio (NR) signal at center frequency $f_c$ that is jammed on its SSB. Here, the SSB jammer permanently transmits uniform noise at center frequency $f_\textnormal{SSB} < f_c$. The bandwidth of this interference is substantially below the bandwidth of the 5G signal. Note that this signal immediately disappears once the SSB jammer is activated, indicating an effective attack.

\subsection{Experimental Setup}
\label{ssec:expset}
\begin{figure}
\centering
    \includegraphics[width=1\hsize]{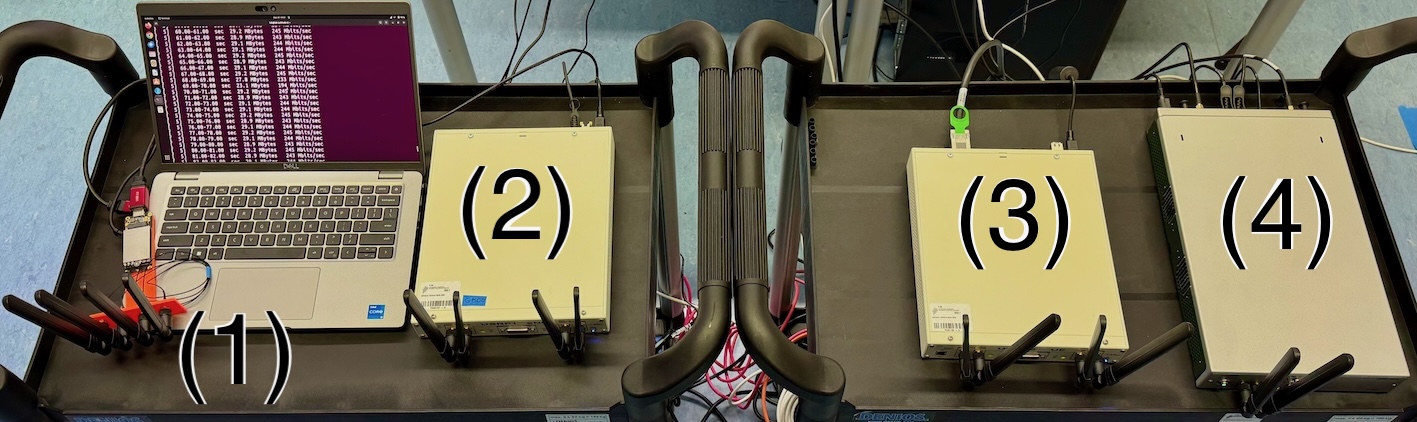}
    \caption{Experimental setup: (1) 5G UE and the SDR frontends for the (2) jammer, (3) watchdog, and (4) gNB. The PCs for (2) to (4) are not shown. The shown distances between the devices are for illustration purposes only. }
    \label{setup}
\end{figure}

Fig. \ref{setup} shows the experimental setup. We are running a private 5G network in the n78 frequency band, at center frequency $f_c = 3750$\,MHz, with $30$\,kHz subcarrier spacing, and $2\times 2$ multiple-input and multiple-output (MIMO). The system operates in time division duplexing (TDD) mode at a signal bandwidth of 100\,MHz or, equivalently, with $|\mathcal{S}|=3333$ subcarriers. The SSB covers $7.2$\,MHz of bandwidth or, equivalently, uses $|\mathcal{S}_\textnormal{SSB}|=240$ subcarriers.

The gNB implements the 5G NR air interface as a Software Defined Radio (SDR) by the software srsRAN 23.10 \cite{srsran} on a personal computer (PC), connected to the universal software radio peripheral (USRP) n300 RF frontend \cite{usrp}. The transmission power of the gNB is 6\,dBm, measured as equivalent isotropic radiated power (EIRP). The UE is a commercial-off-the-shelf 5G modem, namely Quectel RM520N-GL \cite{queltec}. Its maximum transmission power is $23$\,dBm (EIRP). Note that the UE adapts its transmission power, while gNB and jammer transmit at constant power.

The core network functionality is provided by the software Open5GS 2.6.6 \cite{open5gs}, running on the same PC as srsRAN. This setting provides a 5G standalone network and complies with Release 17 of the 5G standard series 38 \cite{3gpp}. 

The jammer and the watchdog run on separate computers, each using one USRP X310 \cite{usrpx} RF frontend. The jammer transmits uniform noise with a duty cycle of $100$\% at center frequency $f_\textnormal{SSB}=3709.92$\,MHz with $7.2$\,MHz of bandwidth and a transmission power of $10$\,dBm. The watchdog permanently records IQ samples over $120$\,MHz bandwidth, including a $10$\,MHz portion of the upper and lower neighboring band, respectively.

The distance between adjacent devices was $1$\,m. Due to this low value, distance is not a relevant factor in our experiments. Without jamming, the above 5G setup reaches an average throughput of 223\,Mbit/s over the air at the transport layer (User Datagram Protocol, UDP). With SSB jamming, neither downlink nor uplink communication is possible.

Dataset creation, training, and measuring the classification time was performed on a single workstation with an Intel Xeon w7-2495X CPU and an NVIDIA RTX A6000 GPU. The operation system was Ubuntu 22.04 LTS with Python 3.9.12 and keras 2.12.0. The GPU was only used for training and is, thus, not utilized during the measurement of classification time.

\subsection{Dataset Creation}
\label{dataset_creation}
A spectrogram is composed of a stack of $n$ arrays, where each array represents the signal's power spectral density (PSD). Each PSD array is obtained by applying an FFT to a window of $m$ IQ samples. We used an FFT rather than Welch’s method \cite{Welsch}, sacrificing precision for computational speed with an average error of $6\%$ for each frequency bin of the PDS array. A spectrogram is then composed as a $n \times m$ matrix.

Based on previous measurement campaigns, we chose a window size of $m=1024$, leading to a frequency resolution of $117$ kHz and a stack of $n=100$ PSDs per spectrogram. This corresponds to a time window of $0.8$ ms.

The resulting spectrogram matrices presented two problems when fed into the training set of the model. First, the power of received radio signals is very low in the linear domain and, thus, typically expressed in the logarithmic domain (decibel). Similarly, we apply the monotonic function $f(x)=-\log x$ to each value of the PSD array at each frequency, to avoid the vanishing gradient problem \cite{vanishing_gradient}. Second, due to approximation errors with very low energy bins, some values of the PSDs turned out to be $0$, which causes computation errors with the log function. Then used a small constant $\epsilon = 10^{-21}$, leading to the applied transformation $f(x)=-\log (x+\epsilon)$. To each sample $i$, a label $y_{i}$ is assigned.

\subsection{Cases and Labeling}
\label{ssec:cases}
The collected spectrogram data is labeled in three cases:
\begin{enumerate}
    \item \textbf{Empty channel, not jammed:} the jammer is not active; gNB periodically transmits beacons but UE is not transmitting
    \item \textbf{Ongoing transmission, not jammed:} the jammer is not active; UE and/or gNB are transmitting and receiving data in TDD mode
    \item \textbf{Jammed:} the jammer is injecting uniform noise into the SSB subchannel; UE and gNB occasionally transmit signals (e.g., beacons, connection requests) but cannot decode received signals
\end{enumerate}
The first two cases are classified as legitimate and labeled with $0$, while the third case is classified as an attack and labeled with $1$.
\newline
The dataset was then uploaded in IEEE DataPort and is available at\cite{dataset}.

\section{Machine Learning Models}
\label{sec:method}

\subsection{Convolutional Neural Network}  \label{sec:cnn}
As spectrograms are two-dimensional, the first model of choice is a CNN whose structure is given in Table \ref{cnn_table}. The objective is to return the correct label associated to each sample at the final layer of the network which consists in one neuron. The chosen loss function is the binary cross-entropy
\begin{equation}
L = -\frac{1}{N}\sum_{i=1}^{N}y_{i}\cdot \log(\tilde{y}_{i})+ (1-y_{i})\cdot \log(1-\tilde{y}_{i}),
\end{equation}
where $\tilde{y}_{i}$ is the prediction of the $i$-th sample. This function can approach infinity even if the prediction error cannot be above $1$, thus, allowing the model to update its weights. 
\begin{table}
  \centering
  \renewcommand{\arraystretch}{1}
  \caption{Structure of the employed CNN}
  \begin{tabular}{c|llr}
    \toprule
    & Layer & Output size & No. of parameters\\
     \midrule
\parbox[t]{2mm}{\multirow{5}{*}{\rotatebox[origin=c]{90}{}}} &     Input              & $100 \times 1024 \times 1$ & $0$\\
     & Convolutional 1    & $49 \times 511 \times 32$ & $320$\\
     & Batch Normalization    & $49 \times 511 \times 32$ & $128$\\
     &Average Pooling   & $24 \times 255 \times 32$ & $0$\\
     & Convolutional 2    & $22 \times 253 \times 64$ & $18496$\\
    & Batch Normalization    & $22 \times 253 \times 64$ & $256$\\
     &Average Pooling   & $11 \times 126 \times 64$ & $0$\\
          & Convolutional 3    & $9 \times 124 \times 128$ & $73856$\\
               &Average Pooling   & $4 \times 62 \times 128$ & $0$\\
    & Batch Normalization    & $4 \times 62 \times 128$ & $512$\\
        & Flattening    & $31744$ & $0$\\
        & Dense    & $64$ & $2031680$\\
        & Dense    & $32$ & $2080$\\
        & Dense    & $16$ & $528$\\
        & Dense    & $1$ & $17$\\

     \bottomrule
  \end{tabular}
  \label{cnn_table}
\end{table}
\newline
Defining $\mathcal C_0$ as the class of \textit{no-jamming} and $\mathcal C_1$ as the class of \textit{jamming}, the jammer detection is performed by test function 
\begin{equation}
\hat{\mathcal C} = \begin{cases}
\mathcal C_0; & \tilde{y}_{i} < \tau \\
\mathcal C_1; & \tilde{y}_{i} \geq \tau,\\
\end{cases}
\end{equation}
on the input sample $\bm{X}$ with a chosen threshold $\tau$. We, thus, measure accuracy as probability of false alarm (FA) and misdetection (MD) defined as
\begin{equation}
P_{\rm FA, C} = {\mathbb P}[\hat{\mathcal C} = \mathcal C_1 |\mathcal C = \mathcal C_0],
\end{equation}
\begin{equation}
P_{\rm MD, C} = {\mathbb P}[\hat{\mathcal C} = \mathcal C_0 |\mathcal C = \mathcal C_1].
\end{equation}

\subsection{Principal Component Analysis}
Using $100 \times 1024$ spectrograms as model input leads to $102400$ dimensions. To avoid the \textit{curse of dimensionality}, we reduce the number of coordinates in two steps. 

First, we average spectrograms over time, reducing each $102400$-dimensional spectrogram to a $1024$-dimensional PSD. This filters out random fluctuations and yields robust PSDs. 

\begin{figure}
\centering
\includegraphics[width=0.9\hsize]{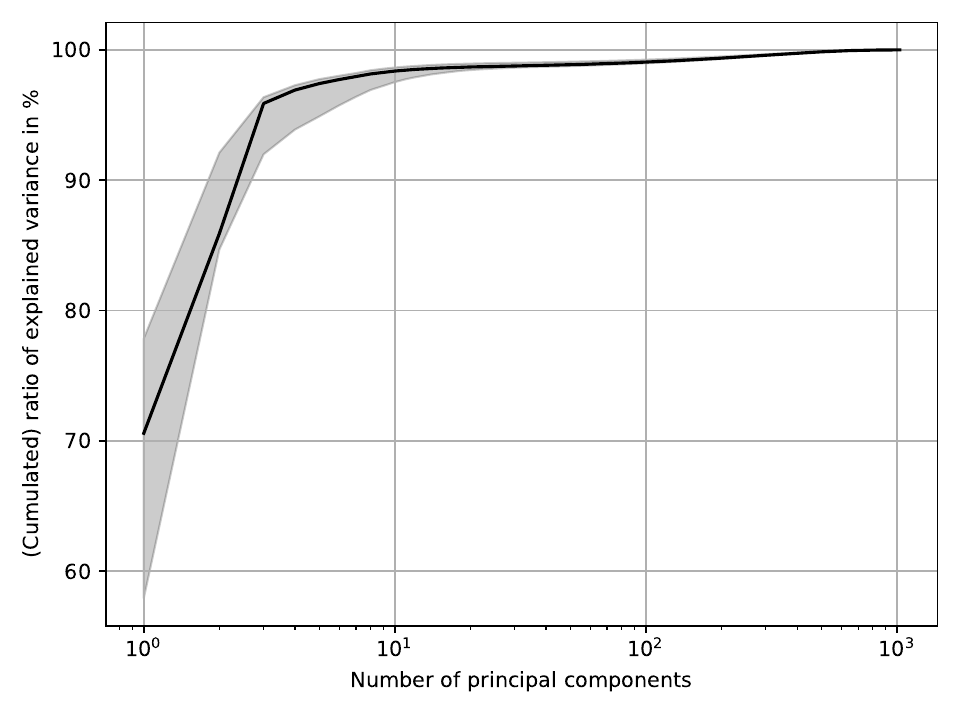}
\caption{Cumulated ratio of explained variance in \% for a varying number of principal components. The $0.01$- and $0.99$-quantiles are based on 100 random datasets and marked in gray}
\label{fig:explained_var}
\end{figure}

Second, we further reduce the dimensions with principal component analysis (PCA). The basic idea of PCA is to change the basis so that the first coordinates (or principal components) contain ``the most information'' about the underlying data. The \textit{ratio of explained variance} measures the information contained in each component and sums up to $1$ when summed over all components. The cumulated ratio of explained variance is displayed in Fig. \ref{fig:explained_var}. It can be seen that a few components already capture most of the information. In particular, 98\%, 98.5\%, and 99\% of the variance are explained by $8$, $13$, and $85$ principal components, respectively. This suggests $8$ components an optimal trade-off between computational complexity and accuracy.

\subsection{$k$-Nearest Neighbors} \label{sec:knn}
For a new sample, KNN searches for the $k$ most similar samples from the training data and makes a prediction based on majority voting. The number of neighbors $k$ is a crucial hyperparameter that needs to be tuned. In Section \ref{sec:results_knn} results for a varying number of neighbors will be discussed. 

KNN is sensitive to the scale of the data. To standardize the 1024-dimensional PSDs, we subtracted their respective means and divided the results by their respective standard deviations. We report results for (i) standardized PSDs and (ii) based on projections of non-standardized PSDs onto the first 8 principal components.

\subsection{Support Vector Machines}
SVMs are a powerful class of machine learning models, which can be used for supervised classification and regression. SVMs operate by constructing a decision hyperplane in the feature space that optimally separates different classes. By employing the kernel trick, support vector machines can also be used
effectively for non-linear tasks. In this work, we employed SVMs with various kernels after dimension reduction with PCA.

\subsection{Classification Time}
To measure how fast each of the above machine learning models can detect a jamming attack, we study classification time. This measure includes the computation time to compute the class and to apply the classification criterion. The time for computing the spectrogram from IQ samples is not included as it is equal for all machine learning models. In the following section, we will compare the cumulative distribution function (CDF) of the classification time obtained over $1000$ trials.

\section{Experimental Results}
\label{sec:results}

\subsection{Convolutional Neural Network}
The training set was composed of $6000$ samples, equally distributed among the three cases described in Sec. \ref{ssec:cases}. Using the same distribution, the validation set was composed of $3000$ samples. This set is used to monitor the validation loss and to stop the training phase after three increases of the parameter. The test set contains $1200$ samples, distributed as the training and validation set.

\begin{figure}
\centering
\includegraphics[width=0.85\hsize]{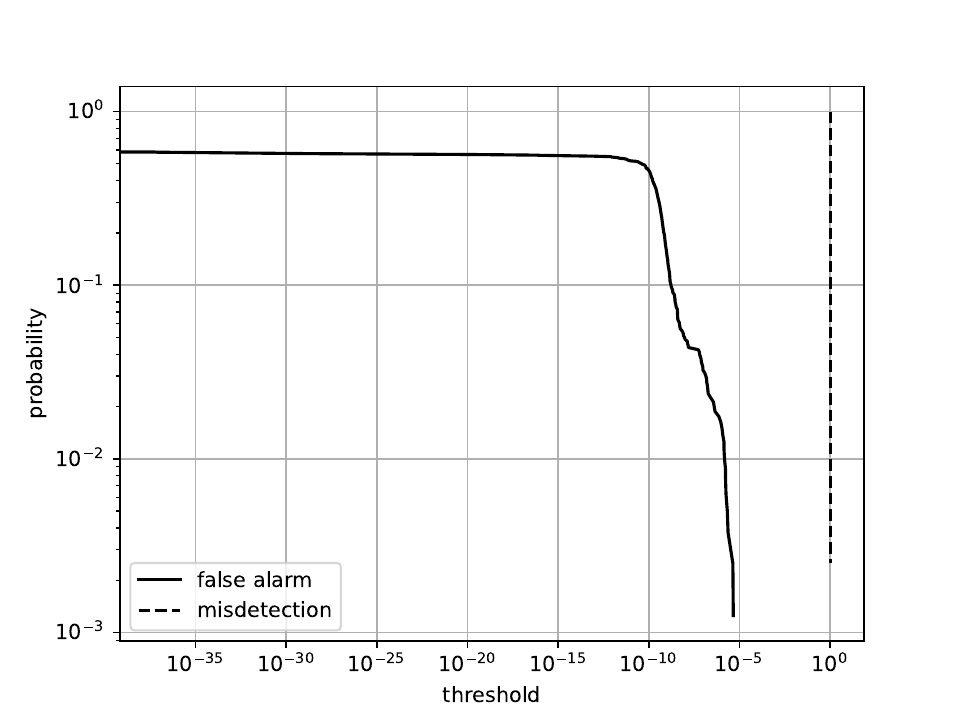}
\caption{FA and MD probabilities as a function of the threshold $\tau\in [0,1]$ (with $y$ axis values normalized to $1$) for the considered jammer}
\label{accuracy_cnn}
\end{figure}

\begin{figure}
\centering
\includegraphics[width=0.85\hsize]{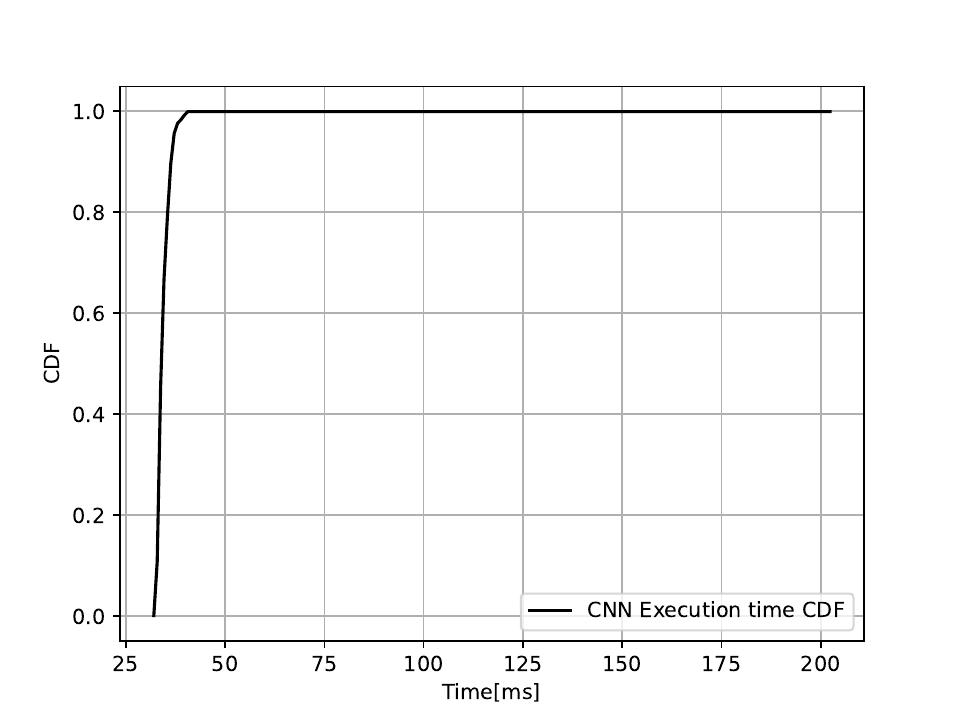}
\caption{CDF of the classification time performed by the CNN with supervised
learning}
\label{time_cnn}
\end{figure}

Fig. \ref{accuracy_cnn} shows that the model perfectly distinguishes between the jamming and no-jamming cases. Fig. \ref{time_cnn} shows the CDF of the classification time for a sample. In $95\%$ of the cases, a classification was achieved within $38$ ms.

To study the performance in a wider range of scenarios while still maintaining experimental control, we reduced the gain of the jammer from 80\,dB to 45\,dB in several steps. Assuming a dominant direct path with distance $d$ and quadratic path loss $d^{-2}$, this corresponds to an increase of the distance between jammer and watchdog from 1 to 56.23\,m. Note that a gain of 80\,dB corresponds to the standard scenario and that all other parameters remain as in Section \ref{ssec:expset}.

\begin{figure}
\centering
\includegraphics[width=0.85\hsize]{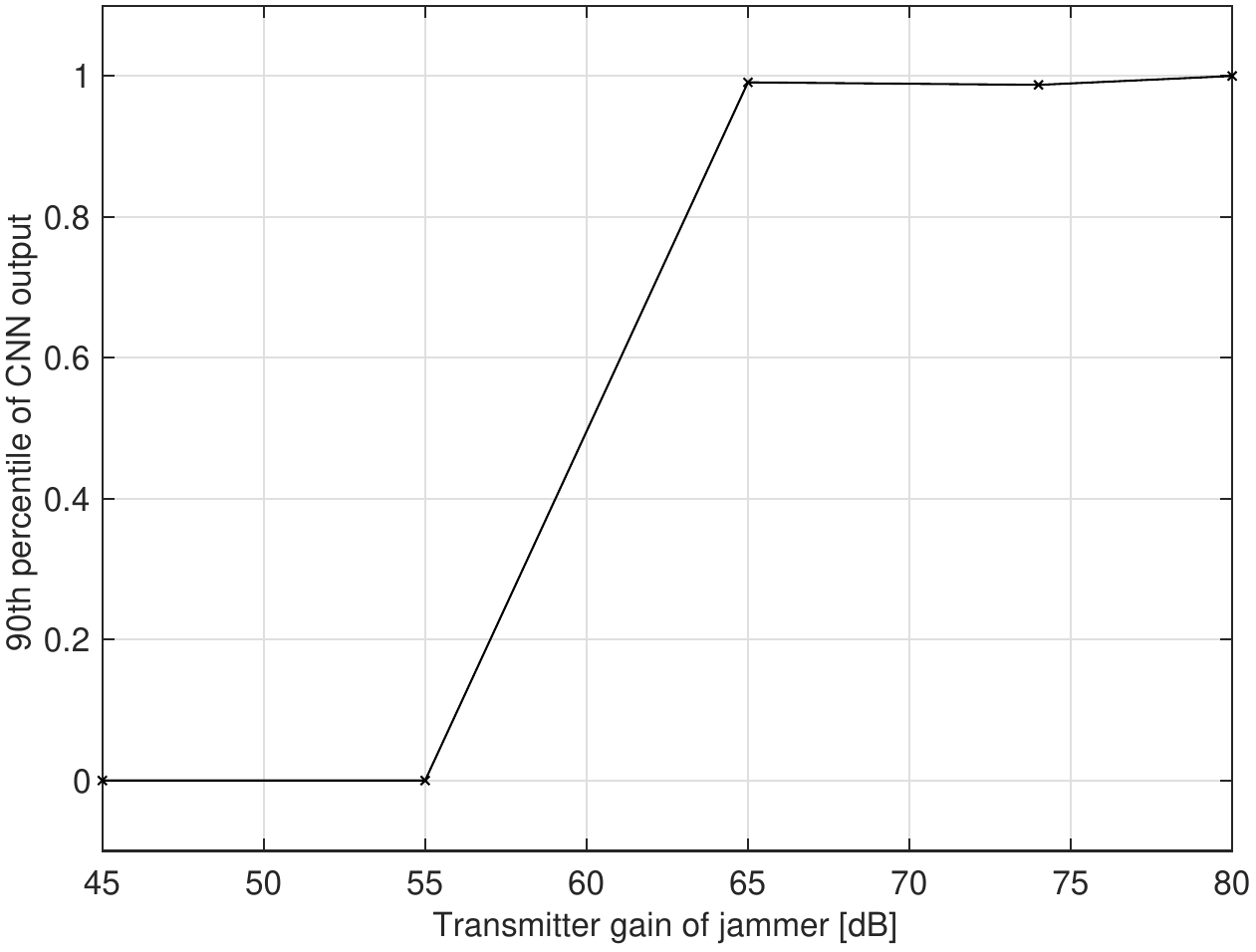}
\caption{90th percentile of the CNN output over transmitter gain of the jammer. With a quadratic path loss model, a gain of 45\,dB corresponds to a jammer-to-watchdog distance of $56.23$\,m, 55\,dB to $17.78$\,m, 65\,dB to $5.65$\,m, 74\,dB to $2$\,m, and 80\,dB to $1$\,m.}
\label{cnn_vargain}
\end{figure}

Fig. \ref{cnn_vargain} shows the resulting accuracy for these experiments as the 90th percentile of the CNN output with an active jammer. In this situation, the correct output of the CNN would be 1 while 0 would be a false negative and no other output values are allowed. Consequently, the closer the shown percentile is to 1, the higher the accuracy. The results show such high accuracy for gains of 65\,dB and above. With above path loss model, this means that jammers within $5.65$\,m distance from the watchdog's antenna can be detected at high accuracy. For larger distances, the accuracy quickly drops to a false negative rate of 100\%.  

Note that, due to the low transmission power of our laboratory equipment, the absolute distance value is not relevant in such geometry discussions. Instead, the ratios between gains (or, equivalently, distances) need to be considered. This means that jammers can still be accurately detected if they exceed the UE-to-basestation distance by a factor of $5.65$. This is an impressive result, although the sudden accuracy drop points to an over-fitted model which can be improved by retraining with various jamming distances.

\subsection{$k$-Nearest Neighbors} \label{sec:results_knn}
For each of the training, validation, and test sets, 1024 samples per class were selected from each of the three cases in Sec. \ref{ssec:cases}. The training, validation, and test accuracy for the settings \emph{(i) standardized PSDs} and \emph{(ii) projections of non-standardized PSDs onto the first 8 principal components} are displayed in Table \ref{knn_table}. The optimal choice of neighbors based on the validation data is $k=8$ for setting (i) and $k=1$ for setting (ii). With an accuracy of approximately $99.98$\%, we can reliably detect jammers. 

\begin{figure}
\centering
\includegraphics[width=0.85\hsize]{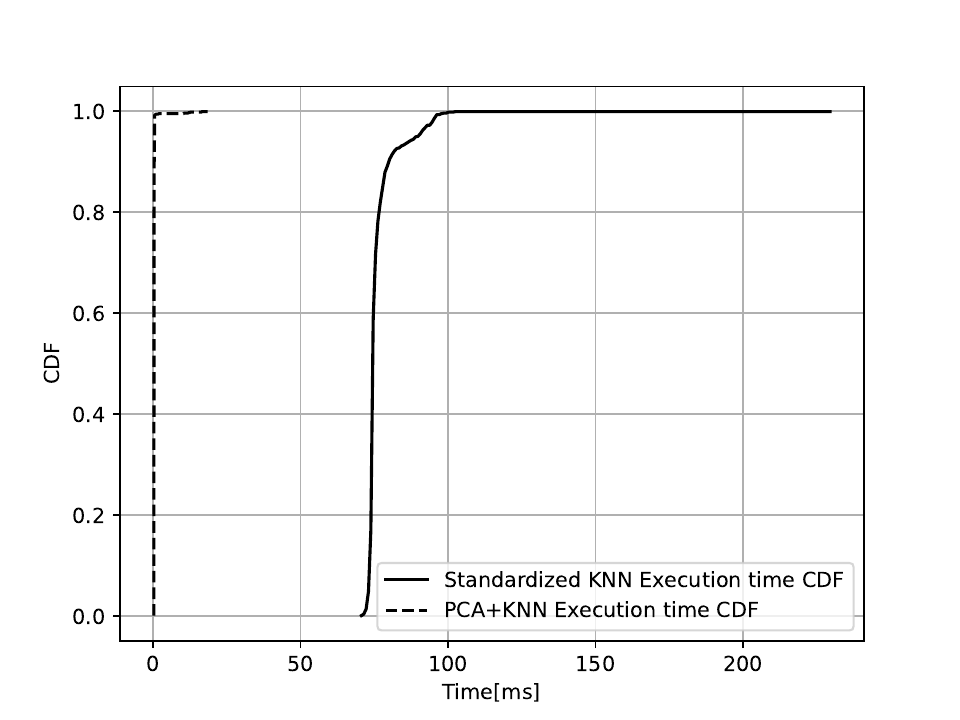}
\caption{CDF of the classification time performed by the KNN with standardization and PCA}
\label{time_knn}
\end{figure}

Fig. \ref{time_knn} shows the CDF of the classification time of the KNN. With standardization, the model classified $95\%$ of the cases within $94$ ms. With PCA, the model performed substantially better with a classification time of $8$ ms for $95\%$ of the cases.

\begin{table}
  \centering
  \renewcommand{\arraystretch}{1}
  \caption{Classification accuracy with $k$-nearest neighbors in \%}
    \begin{tabular}{r|rrr|rrr}
    \toprule
    & \multicolumn{3}{c|}{Standardized} & \multicolumn{3}{c}{PCA (8 comp.)} \\
    $k$ & training & validation & test & training & validation & test \\
    \midrule
	1 & 100.000 & 96.754 & 96.717 & 100.000 & 99.996 & 99.979 \\
	2 & 99.987 & 99.988 & 99.980 & 99.985 & 99.995 & 99.974 \\
	4 & 99.983 & 99.988 & 99.979 & 99.985 & 99.995 & 99.976 \\
	8 & 99.981 & 99.989 & 99.978 & 99.985 & 99.995 & 99.975 \\
	10 & 99.981 & 99.989 & 99.977 & 99.985 & 99.995 & 99.975 \\
	20 & 99.981 & 99.984 & 99.975 & 99.985 & 99.995 & 99.972 \\
	30 & 99.981 & 99.983 & 99.972 & 99.985 & 99.992 & 99.968 \\
	40 & 99.980 & 99.980 & 99.967 & 99.985 & 99.988 & 99.962 \\
	50 & 99.977 & 99.979 & 99.960 & 99.985 & 99.981 & 99.959 \\
    \bottomrule
    \end{tabular}
  \label{knn_table}
\end{table}

%
%
%

\subsection{Support Vector Machines}
Our experiments with SVMs showed that this class of models is well-suited for jamming detection. The overall accuracy on the validation and test dataset significantly depends on the used kernel. We considered linear, polynomial, and radial basis function (RBF) kernels. On the other hand, increasing the number of dimensions does not show a significant effect.

As shown in Table \ref{svm_table}, a large dimensionality reduction with a linear kernel produces models with high confidence scores over the datasets for validation and testing. 

\begin{figure}
\centering
\includegraphics[width=1\hsize]{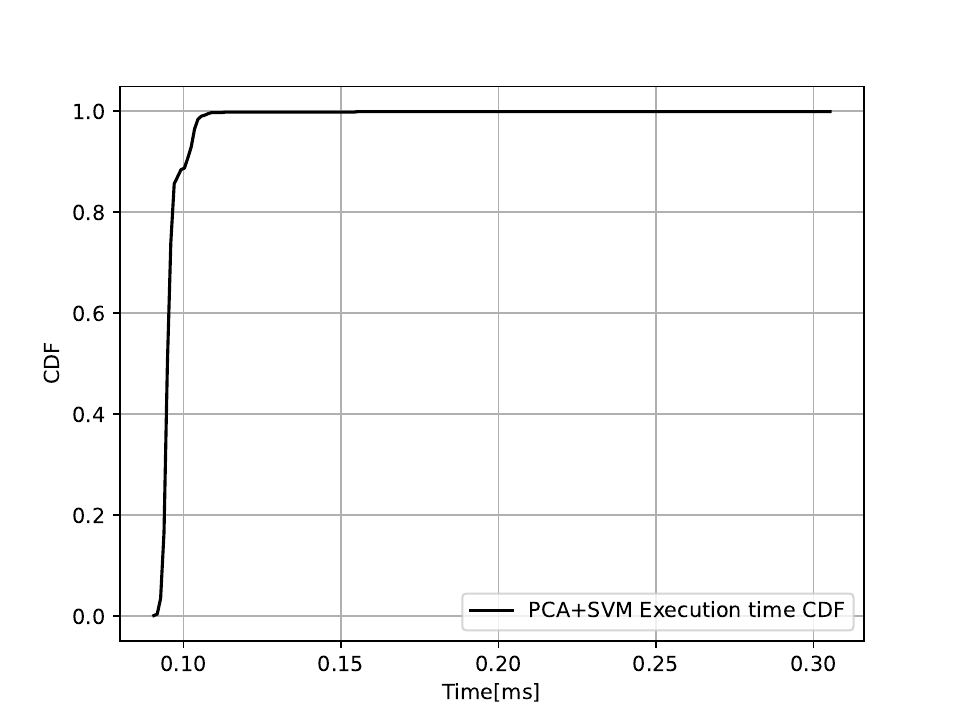}
\caption{CDF of the classification time performed by the SVM with PCA}
\label{time_svm}
\end{figure}

This large dimension reduction highly reduces the computational complexity of the classification. Fig. \ref{time_svm} shows that the SVM model classified 95\% of the cases within $0.11$ ms. Thus, SVMs can detect jammers using substantially smaller computation time than the CNN (Fig. \ref{time_cnn}) and than the KNN without PCA (Fig. \ref{time_knn}). 

\begin{table}
  \centering
  \renewcommand{\arraystretch}{1}
  \caption{Classification accuracy with Support Vector Machines in $\%$}
  \begin{tabular}{llrrr}
    \toprule
    Kernel & Dim & Train Score & Val. Score & Test Score\\
     \midrule
    linear& 8 & 99.31&	95.51&	94.75\\
    linear& 13 & 99.63& 93.52& 93.04 \\
    linear& 85 & 100 &	94.45&	94.21\\
    linear& full dataset & 100&	96.30&	95.88\\
    \midrule
    polynomial& 8 & 98.60&	98.20&	98.01\\
    polynomial& 13 & 98.57&	98.25&	98.14\\
    polynomial& 85 & 98.57&	98.20&	98.14\\
    polynomial& full dataset & 99.99&	98.98&	98.73\\
    \midrule
    rbf& 8 & 98.66&	98.66&	98.45\\
    rbf& 13 & 98.66&	98.74&	98.39\\
    rbf& 85 & 98.63&	98.77&	98.40\\
    rbf& full dataset & 98.71&	98.70	&98.37\\
     \bottomrule
  \end{tabular}
  \label{svm_table}
\end{table}

\section{Conclusion}
\label{sec:concl}  
We have compared three machine learning models to detect an SSB narrowband jammer in 5G networks. Based on data from an experimental 5G network, we studied the computation time and accuracy of these models. 

All three models yielded very high accuracy. CNN correctly classified $100$\% of the jammed cases, followed by test scores of up to $98.98$\% with SVM, and by test scores of up to $99.96$\% with KNN. Thus, in terms of accuracy, either of these methods can be chosen.

Particularly interesting are the results for computation time due to their substantial differences between the machine learning methods. While the CNN required 38 ms to classify 95\% of the cases, KNN needed only 8 ms if combined with PCA. The fasted classification was achieved by SVM with PCA, with a classification time of $0.11$ ms for 95\% of the cases. 

Of further interest is a first impression on detection performance for variable distances. While being limited to the CNN, the results already show that the watchdog does not have to be close to the jammer in order to detect it. The sudden drop for large distances points to an over-fitted model. This, however, can be improved by retraining with various jamming distances.

Motivated by these promising results, we will now study the robustness of the detection accuracy for various wireless network geometries and for multiple jamming signals. 
\balance

\bibliographystyle{IEEEtran}
\bibliography{biblio.bib}
\end{document}